# Designing Zeeman slower for strontium atoms – towards optical atomic clock

Marcin BOBER*, Jerzy ZACHOROWSKI, Wojciech GAWLIK

Marian Smoluchowski Institute of Physics, Jagiellonian University, Reymonta 4, 39-059 Kraków, Poland

*Corresponding author: marcin.bober@uj.edu.pl

We report on design and construction of a Zeeman slower for strontium atoms which will be used in an optical atomic clock experiment. The paper describes briefly required specifications of the device, possible solutions, and concentrates on the chosen design. The magnetic field produced by the built Zeeman slower has been measured and compared with the simulations. The system consisting of an oven and Zeeman slower are designed to produce an atomic beam of $10^{-12}$ s$^{-1}$ flux and final velocity of ~30 m/s.

## 1. Introduction

Strontium atoms, as other alkali earth metals with two electrons in *s* electron shell, are very interesting because of their $J = 0$ ground state. This specific electron structure is the reason for the recent experiments with ultra-cold samples of strontium atoms. The doubly forbidden transition between the singlet ground state $^1S_0$ and the triplet excited $^3P_0$ makes strontium atom an excellent candidate for modern optical atomic clocks [1,2]. Cold strontium atoms can be also used in quantum computations [3], simulations of many-body phenomena [4], and other metrology applications [5]. Recently, Bose- Einstein condensation was achieved in $^{84}$Sr isotope [6,7].

### 1.1. Cooling and trapping atoms

Experiments with ultra-cold strontium, like optical atomic clock, require atoms trapped optically in a magneto-optical trap (MOT) [8]. Atoms can be caught in a MOT if their velocity falls within the capture range determined by the laser beam configuration and the width of the used atomic transition. Typical value of the velocity capture range is tens of m/s. Since the vapor pressure of strontium atoms is low, achieving reasonable loading times into a MOT requires using atomic oven heated to the temperature on the order of 500–600 °C. In these conditions the mean atom velocity is about 500 m/s. Since this is much above the trap velocity limit, atoms have to be slowed down prior to their trapping. For this sake one has to prepare a collimated atomic beam and then to slow down atoms, for example in a Zeeman slower [9]. While, in principle, suitably heated strontium vapor cell can be used for experiments with cold strontium atoms [11], Zeeman slower [10] appears to be the only possible solution for a high performance strontium atomic clock since it assures short MOT loading time and reduces the influence of a black body radiation [12] onto the clock transition.

## 2. Slowing strontium

The transition between the ground state $^1S_0$ and the first excited singlet state $^1P_1$ ($\lambda = 461$ nm, transition with $\Delta m = \pm 1$), can be used both for cooling in MOT and for slowing down atoms in the atomic beam. Relatively high transition width, $\Gamma = 2\pi \times 32$ MHz allows reaching high cooling rate. Spontaneous force from the circularly polarized light beam counter-propagating to the atomic beam can be written as

$$F_s = \frac{\hbar k \Gamma}{2} \frac{S(z)}{1+S(z)+4\left[\delta_0+kv-\mu B(z)/\hbar\right]^2/\Gamma^2}, \qquad (1)$$

where $k$ is the wave vector, $\delta_0$ is the detuning of the laser frequency from the atomic transition frequency, $v$ is the speed of an atom, $S(z)$ is the spatially dependent saturation parameter and $\mu$ is the magnetic moment of the excited Zeeman sublevel used for cooling ($\mu B/\hbar$ is the effective magnetic shift of the transition frequency caused by the field intensity $B$). The maximum force can be obtained from (1) when

$$\delta_0 + kv - \mu B(z)/\hbar = 0. \qquad (2)$$

It is important to realize that the above condition requires adjustment of the local Zeeman shift in pace with the decrease of atomic velocity. The proper shaping of the spatial profile of the magnetic field must assure that possibly all atoms are slowed down to the required end-velocity. Absorption and emission of photons are stochastic processes; therefore some self-regulatory mechanism of the force versus velocity must be applied. The slowing process with maximum force (light in exact resonance with the atomic transition) is unstable in this sense (see Fig. 1). The stability of the slowing process can be improved by using the linear part of the force-versus-velocity dependence, where the slope has its maximum. This means that the assumed force is reduced by some factor $\varepsilon$ relative to the maximum force which results in a longer slowing time and increased length where atoms absorb light and slow down. The optimal value of $\varepsilon$ is 0.75 (this value corresponds to maximum of the derivative of the slowing force versus atomic velocity, see Fig.1) but in a real system, where magnetic field may deviate from the ideal shape, it is desirable to choose $\varepsilon$ slightly smaller. In our set-up $\varepsilon = 0.6$ was chosen. This value gives us moderate slowing efficiency and

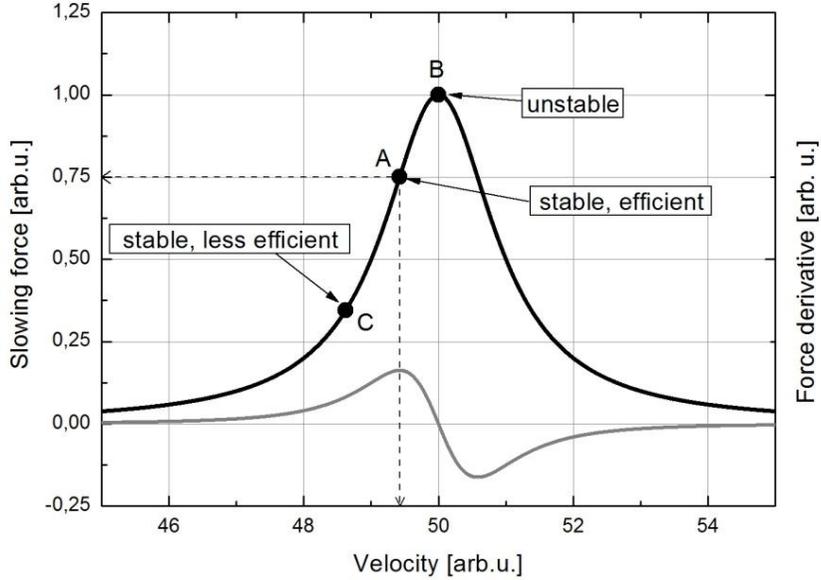

Fig. 1. Dependence of the slowing force (black curve) and its derivative (dark gray curve) on the velocity. Three working points are indicated: in point B the force is maximal, $\varepsilon = 1$, but slowing process is unstable, in A, $\varepsilon = 0.75$, the force versus velocity curve has a maximum slope and slowing process is stable, in C the force and its derivative have lower values.

more stable slowing. Optical force used for calculation of the magnetic field is given by:

$$F_s = \varepsilon \frac{\hbar k \Gamma}{2} \frac{S(z)}{1+S(z)}, \qquad (3)$$

One can calculate the optimal magnetic spatial distribution (for $\sigma^-$ polarized light and $m_J = -1$ excited sublevel) assuming $v(z)$ in the form given by a constant deceleration from an initial velocity $v_0$, under the force of Eq. 3.

$$B(z) = \frac{\hbar}{\mu_B}\left(-\delta_0 - kv(z) + \frac{\Gamma}{2}\sqrt{[1+S(z)]\frac{1-\varepsilon}{\varepsilon}}\right). \qquad (4)$$

The total field range and the length of the slower are determined by the choice of the initial velocity value. A possible offset of the field depends on the laser detuning $\delta_0$. This detuning should be big enough to avoid resonance excitation of atoms outside the Zeeman slower. On the other hand, it would be desirable to make the magnetic field range symmetric around zero so that the absolute value of the field need not be very high. We have chosen $\delta_0 = -2\pi \times 500$ MHz in our case. Atoms with initial velocity higher than $v_0$ will never reach the resonance condition and will not be slowed down. Atoms with initial velocity lower than $v_0$ are not slowed down in the front end of the slower until they enter the region where they reach the resonance condition given by Eq. 4. Once they reach this condition, the magnetic field keep them in resonance and their further evolution is the same as for atoms with initial velocity $v_0$.

## 3. Design

### 3.1. Oven

The atomic oven temperature is an important parameter which determines the final flux of slow atoms at the output of the Zeeman slower. On the one hand, the higher the temperature, the higher the total atomic flux, which is desirable since it reduces the loading time of MOT. On the other hand, the higher the temperature, the wider range of the magnetic field is required in the slower and the longer the slower has to be. We calculated that for an oven with the beam collimation system and temperatures in the range 500–600 °C it is possible to achieve a reasonable flux of $10^{13}$ s$^{-1}$. With the magnetic field range of 600 G it will be then possible to slow atoms with the initial velocity as high as 450 m/s. This means that it will be possible to slow down 35–40% of atoms from the initial flux. Equipping the oven with the collimation stage composed of 8 mm long cylindrical capillaries with 200 µm inside diameter will allow creation of the beam with divergence of 25 mrad. Further collimation will be achieved by two-dimensional laser cooling which, at the same time, will deflect slow atoms to the Zeeman slower region. This will reduce the contamination of our science chamber with the fastest and hottest atoms that cannot be slowed down.

### 3.2. Laser beam

With a moderate laser beam power around 10–25 mW it is possible to slow strontium atoms on a distance of 25–35 cm. To this end we use diode laser (Toptica TA-SHG110). In equation (4) laser intensity dependence is hidden in the saturation parameter $S(z)$. The higher the saturation parameter, the shorter the slower can be. The slower length is an important parameter; it is desirable to shorten it in order to minimize the influence of residual transverse atomic velocities. One also has to remember that during the slowing process atoms are heated which also broadens the transverse velocity distribution.

An interesting improvement can be made by using a beam of a variable intensity for variable transition saturation along the atomic trajectory. This can be created by proper focusing of the laser beam. The variable local saturation parameter has an influence on the required local magnetic field gradient. We used this fact to avoid high gradient at the end of the slower which would be difficult to achieve in a real coil. The saturation parameter was assumed to change from 2.5 at the beginning of the slower to 0.6 at the end. The coil length in such design has increased to 0.305 m from 0.28 m in a design with a constant saturation parameter of 1.5.

### 3.3. Solenoid

The designed magnetic field covered the range of 600 G to slow atoms with the initial velocity of 450 m/s. Assuming higher value of the total field range would increase the fraction of slowed atoms, but at the expense of longer slower which makes the effect of transverse velocities more pronounced. We decided to use a bipolar arrangement of the field, i.e. use two sections producing opposite fields with $|B_{min}| \approx |B_{max}|$. While it is possible to create the required magnetic field with permanent magnets [13], we decided to use a solenoid coil. This offers a possibility to turn off the Zeeman slower magnetic field. Two design ideas were considered to achieve the required field profile: a solenoid with sections of identical windings, but carrying different currents and a solenoid with a profiled winding. The first concept consisting of 12 sections would give us the flexibility in trimming the field profile but, on the other hand, such system would require 12 independent stable current sources. The other concept is less flexible, since the winding is designed for a defined set of parameters, but it requires a single power supply. A program written in Wolfram Mathematica® was used for simulations and calculation of the desired Zeeman coil winding. For given values of the maximum velocity and saturation parameter a theoretical field was calculated with the use of equation (4). Then, using an equation for distribution of magnetic field on axis of a solenoid with given length, thickness and current density, the required coil distribution was simulated. A constant, moderate current value of 3 A was assumed which, together with the wire diameter (1.06 mm), determined the current density. Thickness of the solenoid section with given internal radius of 32 mm was a fitting parameter. After converting the thickness to the discrete number of turns we obtained the spatial distribution of the coil windings. Such a procedure gave relatively thick layer of wire at the very end of the slower (where the magnetic field gradient is the highest). Since this is extremely difficult to realize, we decided to use in this part an independent winding of a thicker wire (diameter 1.7 mm) with high current 9–15 A. This solution gives us also the possibility of fine tuning of the magnetic field profile at the end of the slower with current.

The simulated magnetic field is presented in Fig. 2, together with the measured one. The measured field appears to be in excellent agreement with the simulated one, which proves that simulation of the field worked very well. Fig. 2 shows also the expected magnetic field of the solenoid enclosed in a magnetic shield (see paragraph 3.4) and the theoretical, required field. With the knowledge of the simulated field (with the magnetic shield) it is possible to predict the atom behavior, in particular, the velocity profile. Atom velocity profiles associated with different initial velocities are presented in Fig. 3. For the simulated coil shape the slowing process works for atoms with initial velocity below 450 m/s. Atoms with smaller initial velocity start to be slowed down further downstream within the solenoid only when they reach the resonance condition. All such atoms are slowed down to ~31 m/s.

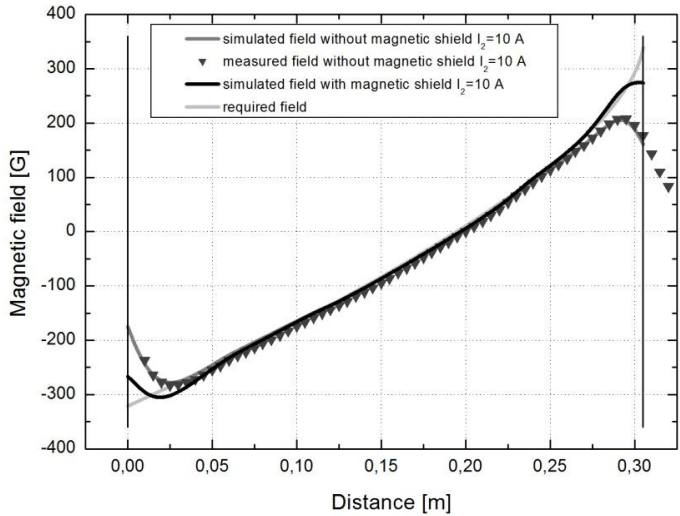

Fig. 2. Magnetic fields created by the coil: the simulated dependence (dark gray) and the measured values (triangles). Black curve corresponds to the simulation of the magnetic field profile of the coil inside a magnetic shield. The required magnetic field profile for the Zeeman slower is represented by the light-gray curve. Vertical lines mark the ends of the solenoid. $I_2$ is current value in the independent section.

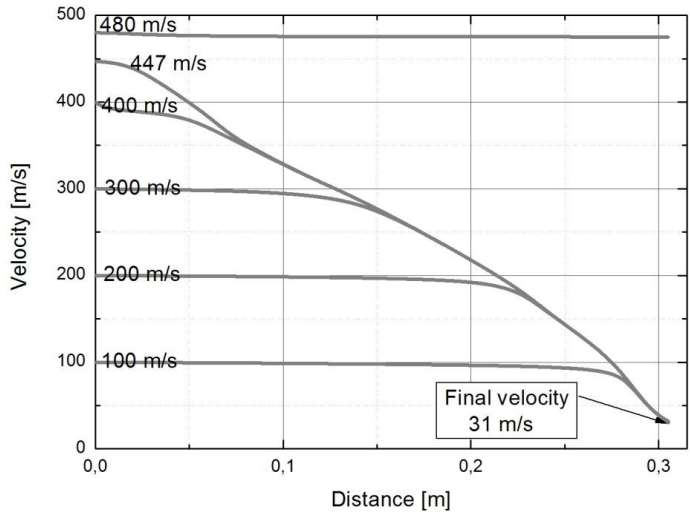

Fig. 3. Atom velocity as a function of distance traveled within the slower plotted for different initial velocities of the atom.

### 3.4. Magnetic shield

Generally, in experiments with optical clocks, atoms need to be free of any perturbation by electric or magnetic fields so it is important to shield atoms from external fields. Application of a magnetic shield offers two advantages. One is that the slower field does not leak outside the shield and does not perturb the MOT phase (the field produced by the Zeeman slower is around 1 G at the position of the MOT in absence of the shield). The second advantage is that the shield increases the magnetic field inside the enclosure. In consequence, higher fields gradient can be created at the end of the Zeeman slower. In our simulations, the effect of a magnetic shield was taken into account as producing a mirror reflection of the field. The advantage of using magnetic shield as a magnetic field mirror is seen in Fig. 2. Two curves present simulated magnetic field with (black) and without (gray) the shield for the same coil. Without the shield it would be necessary to wind more wire turns.

The beginning of the slower is about 7 mm from the magnetic shield wall, which reduces the influence of reflection in the shield wall. The first field maximum is reached a few centimeters inside the solenoid and the capture velocity is decreased. To increase the maximum value and shift its position closer to the beginning of the solenoid we added extra windings which, on the one hand, created higher magnetic field at the very beginning and allowed to capture more atoms with higher velocities but, on the other hand, caused deviation from the required field and decreased the stability margin. Similarly to the entrance to the slower, an extra coil with high current at the other end allows us to increase the magnetic field there, so we are capable to correct the mismatch caused by not ideal shielding by increasing current form 10 to 14 A (Fig. 4).

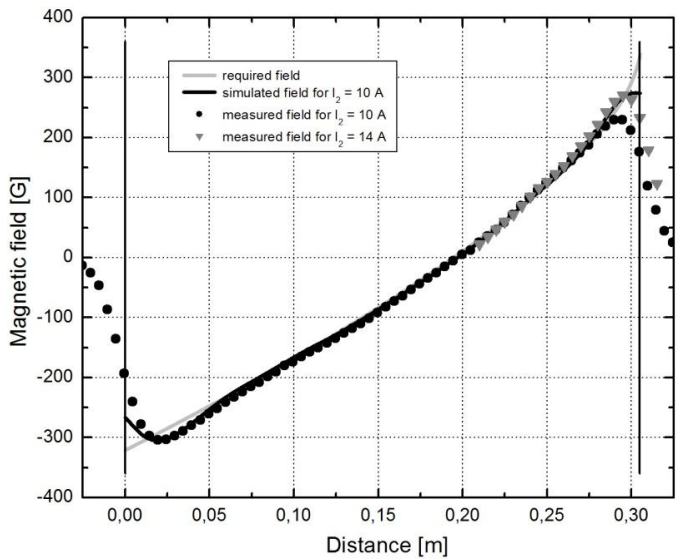

Fig. 4. Required and simulated magnetic field profiles. Black circles correspond to the measurement of field with additional coil section with 10 A current and show significant departure from the simulation at the end of the slower. By increasing the current to 14 A (gray triangles) it is possible to compensate the difference.

Magnetic field measured without the magnetic shield was in excellent agreement with simulations, but the field measured with the shield decreased faster on the edges than in the simulation (see Fig. 4). The best fit could be achieved when we assumed the magnetic-field reflection with efficiency of 75%, but even with such modification the fit was not satisfactory. The remaining discrepancy could be explained by the finite magnetic permeability of the shield material and two holes in the shield wall. We used two-layer magnetic shield with first layer made of pure iron and second of mu-metal. Therefore, any reflection in the shield is not as sharp as in an ideal single layer.

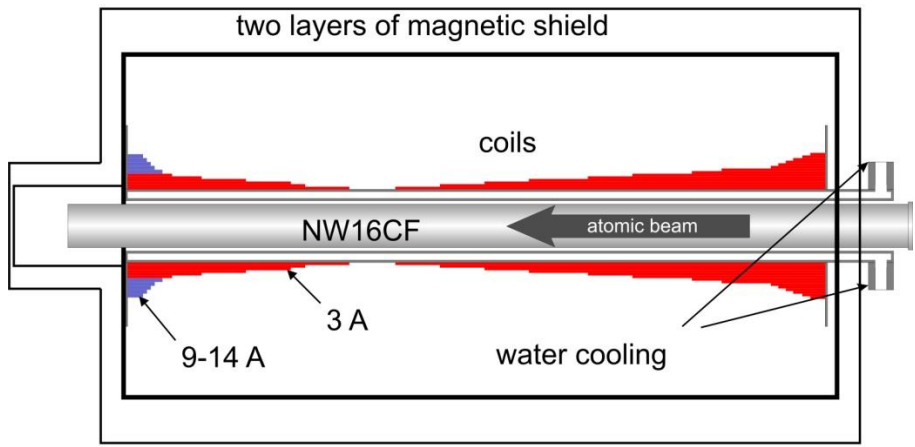

Fig. 5. Scheme of the Zeeman slower. Coil length is 305 mm, inner diameter is 32 mm.

## 3.5. Construction

We decided to apply forced water cooling of the coil. The water-cooling jacket is installed between the coil and the vacuum tube. This increased the inner radius of the winding (additional 4 mm in radius), but only slightly increased the number of solenoid layers required to produce the desired field. For the sake of maintenance we decided to wind the coil on a separate copper holder, rather than directly on the vacuum tube. In Fig. 5 the design idea is shown with the windings and water-cooling layer. The innermost is the vacuum tube NW16CF with outer diameter of 19 mm and knife edge size 21 mm. It is not critical to achieve a perfect magnetic-field profile on the oven side of the slower, where atoms are the fastest. Any deviation of the magnetic field will only change the initial velocity from which atoms will be slowed down, so we decided to shield less efficiently on this side (simple holes instead of extended shielding sleeves used on the MOT side of the slower) and connect water for cooling there.

## 4. Conclusions

The basic principles of Zeeman slowing of atomic beams and specific design for slowing down strontium-atom beam were presented. Due to the focusing of the slowing laser beam and the bipolar magnetic field (-310 to 280 G), the slower of a compact size (coil length 0.305 m) and low power requirements was designed. The atoms of initial velocity below 450 m/s are slowed down to 31 m/s. The Zeeman slower magnetic field has been measured and compared with simulation made before winding.

The slower will be used in the optical clock experiment with strontium atoms. The atoms will be cooled down in a MOT working on the $^1S_0$–$^1P_1$ transition and then further cooled by a laser working on the narrow $^1S_0$–$^3P_1$ intercombination transition (689 nm). Such prepared atomic sample will then be subjected to a destructive measurement of the transition frequency. For reaching practical clock operation it is important to recharge the apparatus with the next cold-atom sample as quickly as possible. Appropriately high flux of slow atoms from the Zeeman slower can reduce the cycle time. Therefore proper design of the Zeeman slower is crucial for efficient functioning of the atomic clock.

*Acknowledgements* – This work was performed as a part of general program of KL FAMO (Ministry of Science and Higher Education grant NN202148933) and was operated within the Foundation for Polish Science Team Programme co-financed by the EU European Regional Development Fund, Operational Program Innovative Economy 2007–2013.

## References

[1] G.K. Campbell, A.D. Ludlow, S. Blatt, J.W. Thomsen, M.J. Martin, M.H.G.de Miranda , T. Zelevinsky , M.M. Boyd, J. Ye, A. Diddams, T.P. Heavner, T.E. Parker, and S.R. Jefferts, *The absolute frequency of the $^{87}Sr$ optical clock transition*, Metrologia **45**, 2008, 539–548
[2] J. Lodewyck, P.G. Westergaard, and P. Lemonde, Nondestructive measurement of the transition probability in a Sr optical lattice clock, Phys. Rev. A **79**, 2009, 061401(R)
[3] A.J. Daley, M.M. Boyd, J. Ye, and P. Zoller, *Quantum computing with alkaline-earth-metal atoms*, Phys. Rev. Lett. **101**, 2008, 170504
[4] M. Hermele, V. Gurarie, and A.M. Rey, Mott insulators of ultracold fermionic alkaline earth atoms: underconstrained magnetism and chiral spin liquid, Phys. Rev. Lett. **103**, 2009, 135301
[5] F. Sorrentino, A. Alberti, G. Ferrari, V.V. Ivanov, N. Poli, M. Schioppo, and G.M. Tino, *Quantum sensor for atom-surface interactions below 10 μm*, Phys. Rev. A **79**, 2009, 013409
[6] S. Stellmer, M.K. Tey, B. Huang, R. Grimm, and F. Schreck, *Bose-Einstein condensation of strontium*, Phys. Rev. Lett. **103**, 2009, 200401

[7] Y.N. Martinez de Escobar, P.G. Mickelson, M. Yan, B.J. DeSalvo, S.B. Nagel, and T.C. Killian, *Bose-Einstein condensation of $^{84}Sr$*, Phys. Rev. Lett. **103**, 2009, 200402

[8] E.L. Raab, M. Prentiss, A. Cable, S. Chu. and D.E. Pritchard, *Trapping of Neutral Sodium Atoms with Radiation Pressure*, Phys. Rev. Lett. **59**, 1987, 2631–2634

[9] W.D. Phillips, H. Metcalf, *Laser Deceleration of an Atomic Beam*, Phys. Rev. Lett. **48**, 1982, 596

[10] I. Courtillot, A. Quessada, R.P. Kovacich, J.-J. Zondy, A. Landragin, A. Clairon, and P. Lemonde, *Efficient cooling and trapping of strontium atoms*, Optics Letters, **28**, 2003, 468–470

[11] X. Xu, T.H. Loftus, J.L. Hall, A. Gallagher, and J. Ye, *Cooling and trapping of atomic strontium*, J. Opt. Soc. Am. B, **20**, 2003, 968–976

[12] L. Hollberg, and J.L. Hall, Measurement of the shift of Rydberg energy levels induced by black body radiation, Phys. Rev. Lett. **53**, 1984, 230–233

[13] Y.B. Ovchinnikov, *A Zeeman slower based on magnetic dipoles*, Opt. Commun. **276**, 2007, 261–267